\documentclass[final,5p,times,twocolumn]{elsarticle}

\usepackage{amssymb}

\usepackage{lineno}

\journal{Nuclear Instruments and Methods A}

\begin{document}

\begin{frontmatter}



\title{Optimisation of CMOS pixel sensors for high performance vertexing and tracking}


\author[A]{J\'er\^ome~Baudot}
\author[A]{Auguste~Besson}
\author[A]{Gilles~Claus}
\author[A]{Wojciech~Dulinski}
\author[A]{Andrei~Dorokhov}
\author[A]{Mathieu~Goffe}
\author[A]{Christine~Hu-Guo}
\author[A]{Levente~Molnar}
\author[A]{Xitzel~Sanchez-Castro}
\author[A]{Serhyi~Senyukov}
\author[A]{Marc~Winter} 
\address[A]{IPHC, Universit\'e de Strasbourg/CNRS, 23 rue du Loess 67037 Strasbourg, France }

\begin{abstract}
  CMOS Pixel Sensors tend to become relevant for a growing spectrum
of charged particle detection instruments. This comes mainly from their
high granularity and low material budget. However, several
potential applications require a higher read-out speed and radiation 
tolerance than those achieved with available devices based on a 0.35~$\mu m$ feature size technology. 
This paper shows preliminary test results of new prototype sensors manufactured in a 0.18~$\mu m$ 
process based on a high resistivity epitaxial layer of sizeable
thickness. Grounded on these observed performances, we discuss a development strategy over the coming years to reach a full scale sensor matching the specifications of the 
upgraded version of the Inner Tracking System (ITS) of the ALICE 
experiment at CERN, for which a sensitive area of up to $\sim$10~m$^2$ 
may be equipped with pixel sensors.

\end{abstract}


\end{frontmatter}


\section{Introduction}
\label{intro}

   CMOS Pixel Sensors (CPS) offer the possibility for subatomic 
physics experiments to address, with unprecedented sensitivity,
 studies requiring an accurate reconstruction of short 
living and low momentum particles. The STAR-PXL vertex detector at RHIC 
\cite{bib:star}, preparing for commissioning, is the first vertex 
detector based on this technology. Its architecture serves as a 
baseline for the inner layers of more demanding devices, such as the CBM Micro-Vertex 
Detector (MVD) \cite{bib:cbm} and the upgraded version of the 
ALICE Inner Tracker System (ITS) \cite{bib:alice}. CPS seem also 
to be well suited to the requirements of vertex detectors at the ILC
\cite{bib:ild} and may open the
possibility to equip large surfaces with fine grained pixels, such as the external layers of inner 
tracking systems, which is an option considered for the future ITS of ALICE \cite{bib:its}.

CPS exploit the industrial technology producing micro-circuits 
equipping common commercial products.
The relevance of CPS for high precision vertexing follows from their 
granularity and material budget going well beyond the LHC standards, 
combined with a capability to cope with high hit rates at moderate 
power dissipation and with a particularly low cost. 
Details on the principle of operation of CPS may be found in \cite{bib:cps}.

Present CPS of the MIMOSA series, developed in a commercial 0.35~$\mu m$ CMOS process, cope with hit rates in the ordre of 10$^6$ particles/cm$^2$/s. 
The limiting factor to reach higher performances required by forthcoming experiments, is now the technology. A 0.18~$\mu m$ CMOS imaging process has become comercially available in recent years, which seems suited to further exploit the potential of CPS. This paper presents laboratory and beam test results of MIMOSA prototypes fabricated in this new technology during 2012.\\

The paper starts (section \ref{secStateOfTheArt}) 
with an overview of the 
state-of-the-art MIMOSA-28 sensor equipping the STAR-PXL 
detector. Next (section \ref{secRequirements}), performance improvements required 
for future detectors are summarised, together with a description 
of the development strategies to match these requirements. 
A description of the first prototype chips (MIMOSA-32) fabricated to explore the performances of the new CMOS process follows.
Preliminary characterisations are described in section \ref{secPerformances}. Finally, section \ref{secConclusion} summarizes the essential results obtained and provides  
an outlook of the major steps leading to a complete 
sensor adapted to the nearest future project: the ALICE-ITS.

\section{State-of-the-art CPS}
\label{secStateOfTheArt}


   The state-of-the art of CMOS pixel sensors is represented by 
the 50~$\mu m$ thin MIMOSA-26 sensor equipping the beam telescope 
of the EU/FP-6 project EUDET \cite{bib:eudet}, and by the MIMOSA-28 \cite{bib:mimo28}
sensors currently assembled on the new vertex detector (called PXL) 
of the STAR experiment at RHIC.

   The architecture of both sensors is based on a column parallel 
read-out with amplification and correlated double sampling (CDS) 
inside each pixel. Each column is terminated with a high precision,
offset compensating, discriminator \cite{bib:discri} and is read 
out in a rolling shutter mode at a frequency of 5~MHz (200 ns/row). 
The discriminators' outputs are processed through an integrated 
zero suppression logic \cite{bib:suze}. 
With its density of 2-3$\times$10$^5$ sensing nodes/cm$^2$, 
this architecture is adapted to a hit rate $>$ 10~$^6$ hits/cm$^2$/s, while dissipating a power below 150~mW/cm$^{2}$. With a pixel pitch of $\sim$ 20~$\mu m$, a single point resolution 
of $\sim$ 3.5~$\mu m$ is regularly achieved with binary charge encoding.

\begin{figure}[h]
\begin{flushleft}
\includegraphics[width=0.4\textwidth]{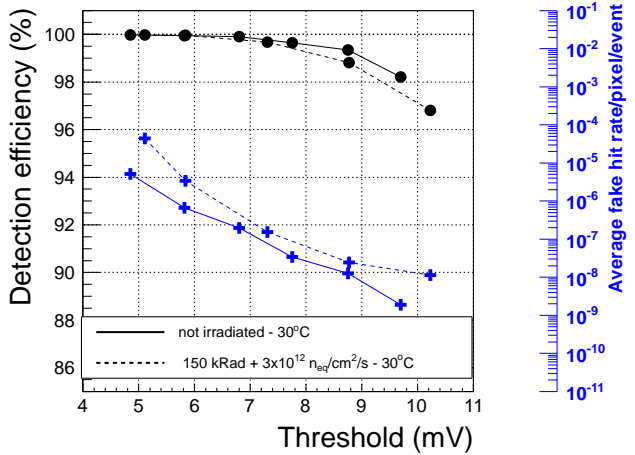}
\caption{Measurement of MIMOSA28 charged particle detection efficiency 
(bullets) and fake hit rate (crosses) as a function of the
threshold values (in mV) of the discriminators implemented in the sensor before (solid lines) and after irradiation (dashed lines). 
}
\label{fig:mimosa28-perfo}
\end{flushleft}
\end{figure}

   The detection performances of MIMOSA-28 were assessed with 
$\sim$ 100~GeV particles at the CERN-SPS. The detection efficiency 
and the fake hit rate (fraction of pixels generating a noise 
fluctuation above threshold) are displayed in 
figure~\ref{fig:mimosa28-perfo} as a function of the discriminator 
threshold, before and after radiation loads corresponding to the 
STAR-PXL specifications at the expected operating temperature of 30$^{\circ}$C. The efficiency is maintained at nearly 100~\%, while the fake hit rate is kept below $10^{-4}$, in ordre to allow the
track reconstruction to remain unaffected by spurious hits due to electronic noise.

 The fact that CPS combine the particle detection and the FEE in 
the same device is a prominent advantage, but turns into a limitation 
as their industrial manufacturing relies on parameters optimised for 
commercial items which may depart substantially from those needed for 
charged particle detection. The high potential of CPS tends therefore 
to be mitigated. CMOS industry has fortunately evolved in a direction 
which allowed CPS to progressively approach their real potential.  

   An important step was achieved when epitaxial layers featuring
a resistivity in excess of 1~k$\Omega \cdot$cm became available for the existing technology. 
SNR values of 30-40 were obtained at room temperature with minimum 
ionising particles, resulting in a tolerance to non-ionising 
radiation damage improved by more than one ordre of magnitude \cite{bib:nonionisingTolerance}.

\section{Performance improvement strategy and new prototypes}
\label{secRequirements}

   Adapting CPS to future applications addresses mainly their 
read-out speed and radiation tolerance. This is illustrated by 
table~\ref{tabRequirements}, which displays the performances of the MIMOSA-28 
sensor equipping the STAR-PXL, and compares them to the requirements of several 
upcoming vertex detectors (ALICE-ITS, CBM-MVD, Super B factories, ILD-VXD) 
which rely on high granularity and low material budget and can 
thus not be equipped with hybrid pixel sensors.

  Among the different components to modify for an improved hit 
rate capability, the pixel dimensions and integrated circuitry 
are the most crucial parameters. Indeed, strategies to accelerate the current column-parallel read-out include: elongating the pixel in one dimension to decrease the number of rows to be read; reading two or four rows simultaneously instead of just one; subdividing the matrix in four to eight sub-areas read in parallel; integrating a discriminator inside the pixel allowing a full digital matrix treatment. A read-out time shortage of up to two ordres of magnitude may actually come out, extrapolating from achieved chip prototyping and simulations.\\
Concerning the radiation impact, an increased resistivity 
of the epitaxial layer, a smaller feature size and a faster 
read-out are expected to enhance the ionising and non-ionising radiation tolerance by at 
least one ordre of magnitude.

\begin{table}[hbp]
\begin{center}
\renewcommand{\arraystretch}{1.1}
\begin{tabular}{|l|ccc|}
\hline 
  Experiment-System & $\sigma_t$ & TID & Fluence \\
				& ($\mu$s)     & (MRad) & (n$_{eq}$/cm$^2$) \\
\hline  
\hline
 STAR-PXL & $\lesssim$ 200 & 0.150 & $3 \times 10^{12}$ \\
\hline
 ALICE-ITS  & 10-30 & 0.700 & $10^{13}$  \\
 CBM-MVD    & 10-30 & $\lesssim$ 10 & $\lesssim 10^{14}$ \\
 ILD-VXD    & $\lesssim$ 10 & ${\cal O}$(0.1) & ${\cal O}(10^{11})$ \\
 Super B factories & $\lesssim$ 2 & 5 &  $5 \times 10^{12}$ \\
\hline
\end{tabular}
\caption{Comparison of the requirements of various vertex detectors, in terms of read-out speed ($\sigma_{t}$) and radiation tolerance related to the total ionising dose (TID) and non-ionising particle fluence.}
\label{tabRequirements}
\end{center}
\end{table}


  It follows from the previous section that accessing a CMOS 
technology of small feature size, is highly desirable. 
But few processes offre the necessary signal detection conditions through
an adequate epitaxial layer. The relatively recent availability 
of the 0.18~$\mu$m CMOS Image Sensor process provided by {\it TowerJazz} has therefore triggered a great 
interest.\\
We stress three features among the most attractive ones of this technology. Six metal layers are available for dense circuitry. Buried P-implants (quadruple well) allow to integrate PMOS transistors inside the pixel, and hence more complex functions, without adding a competitive N-well for charge collection. The process offres an epitaxial layer with resistivity in the range 1 to 5~k$\Omega$.cm.

During late 2011 and 2012 several chips (MIMOSA~32) were manufactured in the aforementioned 0.18~$\mu$m technology, based on a high-resistivity, supposedly 18~$\mu$m thick, 
epitaxial layer. They include numerous small matrices (1.28$\times$0.32~mm$^{2}$ sensitive area each), with various designs: pixel size and form factor, in-pixel amplifier, type of transistors integrated in pixels, etc... Each sub-matrix features the same read-out time of 32~$\mu$s, approaching the expected range of future devices depicted in table~\ref{tabRequirements}. The goals were to compare their charge collection properties, poorly predictable by simulation tools, evaluate various preamplifiers and investigate the radiation tolerance of the technology.

\section{Prototype performance evaluation} 
\label{secPerformances}

\begin{figure}[h]
\centering
\includegraphics[width=0.50\textwidth]{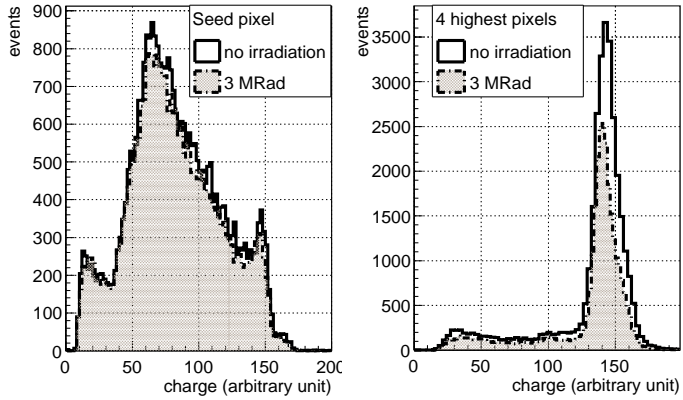}
\caption{Charges collected from $^{55}Fe$ X-rays illumination of a MIMOSA32 pixel including PMOS transistors. Left: seed pixel alone. Right: $2\times2$ cluster. Solid empty histogram: before irradiation, Dotted filled histogram: after a 3~MRad TID.}
\label{fig:mimosa32_charge_xray}
\end{figure}

  The MIMOSA~32 sensors were first studied at room temperature with an $^{55}$Fe source, looking 
at the chip response to the 5.9~keV X-Rays it emits. 
Figure~\ref{fig:mimosa32_charge_xray} displays the distribution of 
the charge collected by $20\time20$~$\mu$m pixels containing a deep P-well hosting two 
P-MOS transistors polarised with the 1.8 V reference voltage of the 
process. On the left, the distribution of the charge collected in the seed pixel of 
the clusters exhibits a small peak at large charge values, which originates from those 
X-Rays impinging the chip in the vicinity of a sensing diode, which 
thus collects its full charge (about 1640 e$^-$). For the most common 
case, the charge is shared among several pixels forming a cluster, 
the seed pixel collecting typically 40-50 \% of the total cluster 
charge. The right panel of the figure shows the charge collected by the set of 4 pixels in a cluster 
with the largest signal. The distribution demonstrates that nearly all the 
cluster charge is concentrated in those 4 pixels, confirming the limited 
diffusion of the charges due to the high resistivity of the epitaxial 
layer.
The figure shows also the distributions measured after an exposure
of the chip to a TID of 3 MRad. No significant degradation is 
observed, indicating that the yearly dose expected in the ITS 
innermost layers ($\lesssim$ 1 MRad) should not modify the sensor
charge collection properties. There is in particular no indication 
of a parasitic charge collection node generated by the deep P-well. 

Beam tests were performed in Summer 2012 at the CERN-SPS 
with 60-120 GeV particles, which allowed evaluating the 
chip detection performances at three different coolant 
temperatures (T$_c$ =15$^{\circ}$C, 20$^{\circ}$C and 30$^{\circ}$C), 
before and after radiation loads of 0.3~MRad combined with 
$3\times10^{12}$~n$_{eq}$/cm$^2$ and 1~MRad combined with 
10$^{13}$~n$_{eq}$/cm$^2$. The study is based on about 10 different 
MIMOSA-32 dies, tested individually on a beam telescope 
composed either of four pairs of microstrip detectors or of six MIMOSA~26 pixel sensors.
 The detection performances (e.g. signal charge collected, pixel noise, 
signal-to-noise ratio (SNR), hit cluster properties, detection 
efficiency) were derived for each temperature-load configuration from a sample of 1500 to 3000 
tracks reconstructed in the beam telescope and traversing 
the pixel array under test.   

Among the different pixel variants investigated, we focus here on comparisons between a square geometry with a 20~$\mu$m pitch and "elongated" geometries based on $20\times33$~$\mu$m$^{2}$ and $20\times40$~$\mu$m$^2$ large pixels. 
The most elongated pixel features a sensing diode area reduced to 9~$\mu$m$^{2}$ instead of 10.9~$\mu$m$^2$ for the other pixel designs. 
For the sake of radiation tolerance and single point 
resolution, the ''elongated'' pixel pattern was staggered. 
The hits reconstructed in the square pixels exhibited 
a typical cluster charge of $\sim$~1100-1200~electrons, 
essentially concentrated in 2 to 4 pixels, about 40-50~\% 
of the charge being collected by the seed pixel.
For the elongated pixels, because of the lower sensing node surface and density, a smaller cluster charge of $\sim$ 800~electrons was observed.

\begin{figure}[h]
\centering
\includegraphics[width=0.40\textwidth]{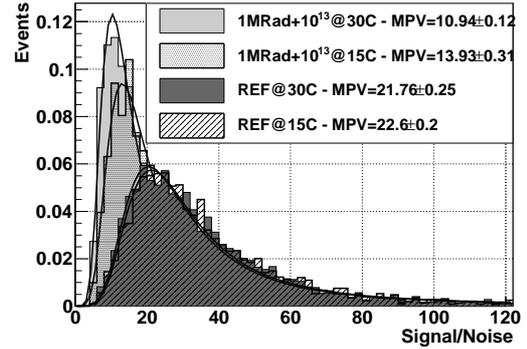}
\caption{MIMOSA32 $20\times40$~$\mu$m$^{2}$ elongated pixel beam test results (60-120~GeV $\pi$): 
SNR distribution for four temperature-radiation load conditions.}
\label{fig:mimosa32-beam-L41-snr}
\end{figure}

The noise values observed reproduce those measured in the 
laboratory ($\lesssim$ 20~e$^-$ENC before irradiation and 
$\lesssim$ 30~e$^-$ENC after irradiation, at 
T$_c$ = 30$^{\circ}$C) for square or elongated pixels without preamplifier. The noise figures for pixels featuring a preamplifier only increase by about 10~\%. 
The SNR values for the $20\times20$~$\mu$m$^{2}$ or $20\times33$~$\mu$m$^{2}$ pixels vary accordingly from 30-35 (MVP) before irradiation to 15-20 for the highest radiation 
load and temperature. The most elongated pixel ($20\times40$~$\mu$m$^{2}$ ) suffers from a reduced SNR, ranging from $\sim$22  to $\sim$11 (before and after irradiation) due to its lowest charge collection efficiency, as depicted by figure \ref{fig:mimosa32-beam-L41-snr}, which displays the SNR for four different load-temperature combinations.\\

\begin{figure}[ht]
\centering
\includegraphics[width=0.45\textwidth]{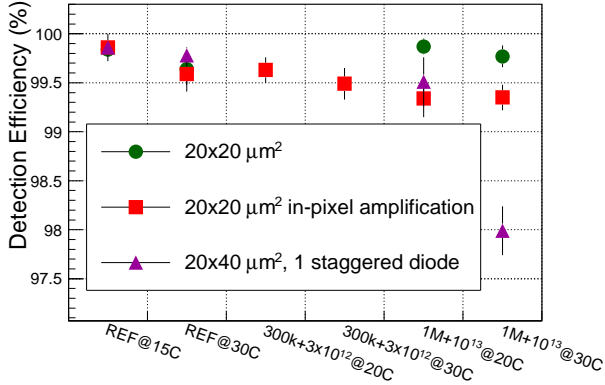}
\caption{MIMOSA32 test beam results: evolution of the detection efficiency with the temperature and radiation load conditions, for three pixel variants.}
\label{fig:mimosa32-eff}
\end{figure}

The detection efficiency shown in figure \ref{fig:mimosa32-eff}, which is about 100~\% before 
irradiation, remains nearly unchanged within 0.5~\% after irradiation for all pixel variants, but the most elongated pixel efficiency is deprecated to 98~\% for the highest radiation load at 30$^{\circ}$C coolant temperature, which has to be reduced at 15$^{\circ}$C to recover a 99.5\% performance. These results demonstrate that a SNR figure around 15 still allows an excellent detection efficiency.

An additional step was performed in the offline beam data analysis, allowing to predict the behaviour of a real size binary output sensor. In the case of the square pixel featuring a preamplifier, a single discriminating threshold was applied to all pixels, and the clustering subsequently operated on a matrix of 0 and 1.

\begin{figure}[ht]
\centering
\includegraphics[width=0.50\textwidth]{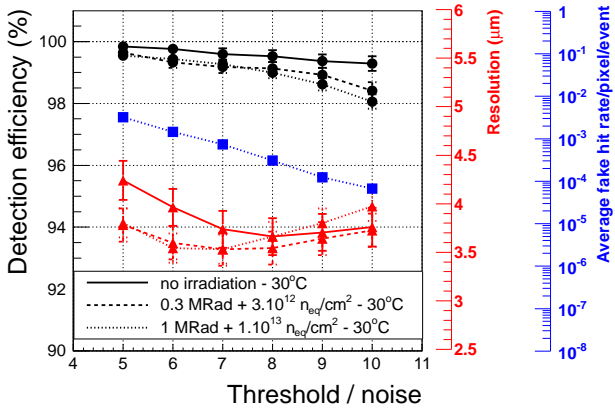}
\caption{Expected performances of a square pixel ($20\times20$~$\mu$m$^{2}$) integrating a preamplification, with emulated binary output from test beam results. The detection efficiency (black bullets), the fake rate (blue squares) and the spatial resolution (red triangles) evolutions with the discriminator threshold (in units of noise) are displayed for three different radiation loads and at the highest coolant temperature of 30$^{\circ}$C.}
\label{fig:mimosa32-binary}
\end{figure}

The detection efficiency and spatial resolution obtained from these emulated digital data are displayed in figure \ref{fig:mimosa32-binary} before and after irradiation for 30$^{\circ}$C coolant temperature. A clear plateau for $>$99~\% efficiency is observed for all radiation loads, ensuring a substantial operating range for the threshold.

The spatial resolution depends on the threshold value, since the latter impacts the cluster pixel multiplicity. The resolution reaches a minimum at 3.5~$\mu$m well inline with expectation for a 20~$\mu$m pitch, binary output pixel. 
 
The observed fake hit rate is significantly higher than measured with sensors fabricated in a 0.35~$\mu m$ process \cite{bib:mimo28}. A more detailed analysis indicated that the noise excess is due to a small fraction of pixels exhibiting RTS (Random Telegraph Signal) like noise \cite{bib:RTS}. The next step of the R\&D is expected to mitigate this effect and downscale the fake rate to an acceptable level by optimising the in-pixel transistors' geometry.

\section{Summary and outlook} 
\label{secConclusion}

 The performances observed with CPS prototypes fabricated in the new $0.18$~$\mu$m technology validate several basic elements needed to proceed forward. First of all the collection properties of the sensitive volume used in the process were shown to match the requirements of the targeted projects.
 
 Accounting for the modest optimisation of the single-diode elongated pixel design and for the high irradiation load and operating temperature considered, the SNR observed are satisfactory and allow taking the $22\times33$~$\mu$m$^{2}$ staggered pixel geometry as a baseline for the ALICE-ITS proposed sensor. 
 
 Finally, a first in-pixel pre-amplifier circuit providing a satisfactory detection efficiency emerged from the various ones tested. The radiation tolerance of the technology was shown to comply with the ALICE requirements.\\

The next step of the development aims at producing the nearly complete
read-out architecture of the final sensor, as well as to optimise the
in-pixel circuitry and mitigate the pixel noise.

The read-out architecture will be investigated with prototypes
representing about 1/4 of the final sensor. The read-out circuitry
encompassing the pixel array and the discriminators will be developed
with one prototype, while the sparse data scan circuitry will
be implemented in a dedicated chip. Two different approaches will
be studied, both based on a rolling shutter where two rows are
read out simultaneously. For one of them, the pixels are grouped in
columns ended with a discriminator, while for the other each pixel
incorporates its own discriminator. The latter case allows for a
twice faster read-out and a twice lower power consumption. The final
- complete - sensor is expected to reach $\sim$~30~$\mu s$ (respectively 15~$\mu s$) read-out time and $\sim$ 300~mW/cm$^2$ (respectively $<$ 200~mW/cm$^2$) if based on end-of-column discriminators (respectively in-pixel discriminators).

  The optimisation of the pixel design is addressed in several additional prototypes.
One of them concentrates on the in-pixel preamplification and correlated
double sampling, a second one addresses the noise reduction and a third
one explores various charge sensing system designs. This latter chip also
features pixels of various dimensions allowing to explore and optimise the
pixel performances as a function of their size.\\
These chips were all submitted to foundry in February-March 2013 and
will be manufactured on wafers featuring epitaxial layers of various
thicknesses and resistivities. Their test through the year 2013 will
lead to a full scale prototype featuring a 1~cm$^2$ sensitive area,
to be fabricated in 2014. It is expected to be the last step before the
the final sensor pre-production, foreseen in 2015.



\begin{thebibliography}{00}

\bibitem{bib:star} L.~Greiner {\it et al.}, 
Nucl.~Instr.~Meth. {\bf A 650} (2011) 68-72.

\bibitem{bib:cbm} M.~Deveaux {\it et al.},  
 PoS {\bf VERTEX2008} (2008) 028.

\bibitem{bib:alice} L.~Musa {\it et al.},  
CERN preprint, March 2012.

\bibitem{bib:ild} the ILD collaboration, 
note DESY 2009-87 and KEK 2009-6, 2009.

\bibitem{bib:its} S.~Senyukov,  
preprint  arXiv:1304.1306, April 2013.

\bibitem{bib:cps} R.~Turchetta {\it et al.}, 
Nucl. Instr. Meth. {\bf A 458} (2001) 677-689.

\bibitem{bib:eudet} the EUDET Consortium, 
Preprint arXiv:1201.4657, January 2012.

\bibitem{bib:mimo28} I.~Valin {\it et al.}, 
JINST {\bf 7} (2012) C01102.

\bibitem{bib:discri} Y.~Degerli {\it et al.},
 Nucl.~Instr.~Meth. {\bf A 602} (2009) 461.

\bibitem{bib:suze} A.~Himmi {\it et al.},
Proceedings TWEPP'09, 2009, September 21-25, Paris.

\bibitem{bib:nonionisingTolerance} A.~Dorokhov {\it et al.},
 Nucl.~Instr.~Meth. {\bf A 624} (2010) 432-436.

\bibitem{bib:RTS} J.~Janesick {\it et al.}, 
Proc. SPIE {\bf 6276} (2006).

\end{thebibliography}
\end{document}